%% file: two_species.tex
\documentclass[aps, pra, twocolumn, amsmath, amssymb, longbibliography, superscriptaddress, floatfix, final]{revtex4-2}

\usepackage{graphicx, dcolumn, hhline, bm, amsmath, amsfonts, amssymb, color, tabularx, comment}

\usepackage[caption=false]{subfig}

\usepackage[%
  pdftex,
  breaklinks=true,
  colorlinks=true,
  linkcolor=blue,
  urlcolor=blue,
  citecolor=blue
]{hyperref}

\newcommand{\Ch}[1]{\color{black} #1 \color{black}}

\begin{document}
\title{\Ch{Modelling dielectric loss in superconducting resonators}: Evidence for \Ch{interacting} atomic two-level systems at the Nb/oxide interface}

\author{Noah Gorgichuk}
\affiliation{Department of Physics and Astronomy, University of Victoria, Victoria, British Columbia V8W 2Y2, Canada}
\affiliation{Centre for Advanced Materials and Related Technology, University of Victoria, Victoria, British Columbia V8W 2Y2, Canada}

\author{Tobias Junginger}
\affiliation{Department of Physics and Astronomy, University of Victoria, Victoria, British Columbia V8W 2Y2, Canada}
\affiliation{Centre for Advanced Materials and Related Technology, University of Victoria, Victoria, British Columbia V8W 2Y2, Canada}
\affiliation{TRIUMF, 4004 Wesbrook Mall, Vancouver, BC, V6T 2A3, Canada}

\author{Rog\'{e}rio \surname{de Sousa}}
\affiliation{Department of Physics and Astronomy, University of Victoria, Victoria, British Columbia V8W 2Y2, Canada}
\affiliation{Centre for Advanced Materials and Related Technology, University of Victoria, Victoria, British Columbia V8W 2Y2, Canada}
\date{\today}

\begin{abstract}
While several experiments claim that two-level system (TLS) defects in amorphous \Ch{surfaces/interfaces}  are responsible for energy relaxation in superconducting resonators and qubits, \Ch{none can} provide quantitative explanation of their data in terms of the conventional noninteracting TLS model. \Ch{Here a model that interpolates between the interacting and noninteracting TLS loss tangent is proposed to perform numerical analysis of experimental data and extract information about TLS parameters and their distribution. As a proof of principle, the model is applied to TESLA cavities that contain only a single lossy material in their interior, the niobium/niobium oxide interface.
The best fits show interacting TLSs with a sharp modulus of electric dipole moment for both thin ($5$~nm) and thick ($100$~nm) oxides, indicating that the TLSs are ``atomic'' instead of ``glassy''. The proposed method can be applied to other devices with multiple material interfaces and substrates, with the goal of elucidating the nature of TLSs causing energy loss in resonators and qubits}.
\end{abstract}

\maketitle

\section{Introduction}

Quantum devices based on superconducting resonators have become one of the most promising architectures for scalable quantum computing \cite{Ball2021, Krantz2019}. Specifically, the use of superconducting radio frequency cavities and qubits have evolved into various promising devices because of their ability to achieve high coherence for quantum states. Maintaining coherence for quantum states is crucial such that quantum information processing can be achieved. However, coherence times for superconducting qubits and resonators are limited in the low energy regime by energy dissipation due to photon loss into two-level systems (TLSs) \cite{Martinis2005, Wang2009, deGraaf2018, Muller2019, McRae2020}, that are most notably present at surfaces/interfaces \cite{Wang2015}. There is currently widespread activity in understanding the microscopic structure of TLSs \cite{Grabovskij2012} and how to remove them with surface passivation techniques \cite{Megrant2012, Earnest2018, deGraaf2018}. In \cite{Diniz2020} it was predicted that if surfaces and interfaces can be made free of TLSs (i.e. if they can be made purely crystalline or epitaxial), their intrinsic photon loss tangent will be reduced by a factor ranging from  $10$ to $10^{4}$ depending on the particular material used. As a result the energy relaxation time of superconducting qubits may reach over $10^{4}$~$\mu$s, putting the architecture above the error correction threshold \cite{Andersen2020}.

Energy losses from TLSs arise from the coupling between the TLS electric dipole moment and the electric field produced by the device. The conventional \emph{noninteracting model} predicts steady-state energy loss described by the loss tangent \cite{VonSchickfus1977, Skacel2015}
\begin{equation}
    \tan{\left(\delta_{{\rm nonint}}\right)}=\frac{\pi p^{2}\rho}{3\epsilon}\frac{\tanh{\left(\frac{\hbar\Omega_0}{2k_BT}\right)}}{\sqrt{1+\left(\frac{E}{E_c}\right)^{2}}},
\label{conv_tan}
\end{equation}
where $p=|\bm{p}|$ is the modulus of the electric dipole moment of the TLS, which is averaged over all possible directions leading to the factor of $3$ in the denominator, and $\epsilon$ is the dielectric constant for the material where the TLS is embedded. The quantity $\rho$ is an energy-volume density (dimensions of $({\rm energy} \times {\rm volume})^{-1}$) for TLSs with energy splitting equal to $\hbar \Omega_0$, where $\Omega_0$ is the resonance frequency of the cavity; $E$ is the electric field at the location of the TLS, and $E_c$ is the characteristic electric field for \emph{saturation}.

Saturation happens when $E\gg E_c$, the number of excited TLSs is maximized (equal to 50\% of the total) so the amount of energy flowing from TLS to photons is equal to the energy flowing from photons to TLS, making the loss tangent equal to zero. It can be shown that
\begin{equation}
   E_c=\sqrt{\frac{3}{2}}\frac{\hbar}{p} \frac{1}{\sqrt{T_1T_{2}}},
   \label{eqnEc}
\end{equation}
where $T_1$ and $T_{2}$ are TLS energy relaxation and homogeneous broadening (coherence) times, respectively, \Ch{due to interaction between TLSs and phonons. Phonons act as a bottleneck for energy dissipation, providing an upper limit for the power than can leak out of the TLS-photon system}. Therefore, measuring the loss tangent as a function of cavity electric field (or input power) provides information on TLS properties. 

\Ch{However, experiments show that loss from surfaces/interfaces have much weaker $E$-field dependence and can not be fit with the noninteracting model (\ref{conv_tan}), even when the geometric dependence of the applied electric field $E$ is accounted for \cite{Wang2009, Macha2010, Wisbey2010, Burnett2016, deGraaf2018, Verjauw2021}. This happens in spite of the fact that measurements of the temperature-dependent resonance shift $\Delta \Omega_0/\Omega_0$ in similar resonators is closely fitted to the corresponding noninteracting-TLS expression \cite{Gao2008, Wisbey2010}}.

\Ch{The usual approach to obtain good fits is to replace the denominator in Eq.~(\ref{conv_tan}) by $\left[1+(E/E_c)^{2}\right]^{\beta}$, and let $\beta$ be a free fitting parameter \cite{Wang2009, Macha2010, Sage2011, Burnett2016, Romanenko2017, Verjauw2021, Altoe2022}. Historically, $\beta$ was introduced to account for the inability of calculating nonuniform $E$ fields in devices with nm-$\mu$m-mm dimensions \cite{Wang2009}, but later $\beta<1/2$ became a proxy for the impact of TLS-TLS interaction on loss \cite{Burnett2016}. Authors report a wide range of best-fit $\beta$ values. For example, \cite{Romanenko2017} shows best fit $\beta=0.25-0.42$ for 13 different samples of niobium, while \cite{Burnett2016} obtained $\beta=0.15$ at different temperatures,
and Fig.~9(e,f) of \cite{Verjauw2021} shows a wide distribution of $0<\beta\leq 1/2$ for identical Nb samples as a function of oxide regrowth. 

TLSs interact through long-range dipole-dipole coupling, due to both electrical and phonon-mediated mechanisms \cite{Enss2005}. This leads to spectral diffusion of the TLS energy level splitting, allowing the photon energy absorbed from one TLS to spread out to other TLSs, circumventing the phonon bottleneck. While all authors seem to agree on the importance of spectral diffusion, there is currently no consensus on how Eq.~(\ref{conv_tan}) gets modified in the presence of interaction. 

Burin and Maksymov \cite{Burin2018} predicted $\beta=1$, in clear disagreement with experiments, motivating the questioning of the assumptions of the standard TLS model. 
In contrast, Faoro and Ioffe \cite{Faoro2012} predicted a logarithmic dependence 
\begin{equation}
    \tan{\left(\delta_{{\rm int}}\right)} = \frac{\pi p^{2}\rho}{3\epsilon}\frac{\tanh{\left(\frac{\hbar\Omega_0}{2k_BT}\right)}}{\ln{(\xi)}}
    \ln{\left(\frac{\xi E_c}{E}\right)},
\label{int_tan}
\end{equation}
with dimensionless ``spectral diffusion'' parameter
\begin{equation}
    \xi = \gamma_{{\rm max}}\sqrt{T_1T_2},
\label{xi}
\end{equation}
describing the ratio between upper and lower frequency cutoffs for spectral diffusion (TLS frequency fluctuating between $\Omega_0 -\gamma_{{\rm max}}$ and $\Omega_0 +\gamma_{{\rm max}}$). The theory in \cite{Faoro2012} focused on asymptotic behaviours, so that Eq.~(\ref{int_tan}) holds provided that two conditions are satisfied: $\xi \gg 1$ and $E_c\ll E\ll \xi E_c$. When at least one of these conditions is not satisfied, the noninteracting result Eq.~(\ref{conv_tan}) was shown to hold \cite{Faoro2012}.  

We are aware of only one attempt to fit Eq.~(\ref{int_tan}) to experimental data \cite{deGraaf2018}; but due to the logarithmic singularity the fit had to be done for a specific value of $E$ and then extended to other regimes.} 

In contrast, there was no attempt to fit statistical distributions for $p$ and $E_c$ (multiple species of TLSs) to the data as is commonly assumed in theories of amorphous materials.

The large spread in observed values of $\beta$ for similar niobium samples raises the question: Can the (non)interacting TLS model (with $\beta=1/2$ or logarithmic dependence) provide the best description under the additional assumption that other TLS parameters such as $p$ and $E_c$ are statistically distributed in amorphous materials? Such a distribution would also slow down saturation, in the same way that $\beta<1/2$ does. 

Answering this question may also shed light on the microscopic structure of TLSs causing loss. There are two known types of TLSs \cite{Enss2005}: ``Atomic TLSs'' occur when impurities tunnel between pairs of equivalent sites. A notable example is Nb:O,H \cite{Wipf1984}, where interstitial O creates a bistable trap for H. Only two locations for H are stable for a given interstitial O, leading to octahedral averaging over the 6 possible directions for $\bm{p}$. \Ch{Such average gives the same result as the angular average used to obtain Eqs.~(\ref{conv_tan})~and~(\ref{int_tan}). Hence the atomic TLS is characterized by a sharply defined $p$}. The other type is the ``glassy TLS'' realized by bistable configurations involving several atoms in an amorphous lattice \cite{Grigera2003, Belli2020}. The wide variation in glassy TLS morphology indicates a broad distribution for $p$, with implications for dielectric loss \cite{Hung2022}. 

Here the question of whether a broad distribution of $p$ is required to explain dielectric loss is investigated by proposing alternative data analysis of experimental data for TLS saturation. \Ch{We propose a model that continuously interpolate between Eqs.~(\ref{conv_tan})~and~(\ref{int_tan}), in order to properly capture the role of spectral diffusion in actual experimental data. As a proof of principle the method is applied to fit} the experimental data of Romanenko and Schuster \cite{Romanenko2017}, who measured quality factor $Q$ in three-dimensional TESLA cavities made of high-quality niobium (in contrast to the artificially doped samples of \cite{Wipf1984}). 
TESLA cavities made of Nb are known to achieve record-high resonance quality factors $Q>10^{11}$, and as a result variations of accelerating cavities adapted to quantum information processing are now under consideration \cite{Kutsaev2020}. 
TESLA cavities \cite{Aune2000} have remarkable structural/materials simplicity when compared to the two-dimensional superconducting devices used in quantum hardware \cite{Woods2019}.
For example, they contain only \emph{one lossy material}, the Nb/oxide interface, \Ch{and no substrate.} 


\Ch{The Nb/oxide interface was shown to contain a thin ($\lesssim 1$~nm) amorphous NbO$_x$ layer covered by crystalline niobium pentoxide Nb$_2$O$_5$, depending on the type of surface treatment \cite{Delheusy2008,Antoine2012, Verjauw2021,Altoe2022}}. It was shown previously that TLSs are present in both types of oxide \cite{Verjauw2021}.

\section{Proof of principle: Numerical modelling of loss in a TESLA cavity}

The resonance quality factor $Q$ measured in experiments is given by $1/Q=\sum_i\tan{(\delta_i)}\times f_i$, where $\tan{(\delta_i)}$ is the loss tangent for a certain region $i$ of the device, and $f_i$ is the corresponding participation ratio \cite{Wang2015}. Participation ratio $f_i\leq 1$ is defined as the fraction of electric energy stored in the dielectric volume $i$, normalized by the total electric energy in the device. 

\begin{figure}[!b]
    \centering
    \includegraphics[width=1\linewidth]{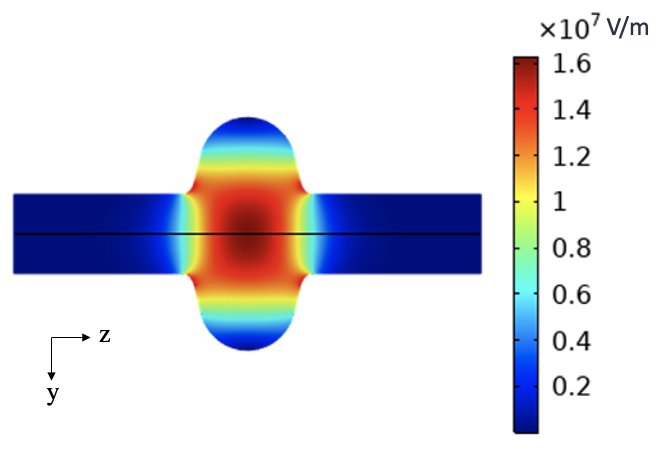}
    \caption{Cross-section of the electric field amplitude distribution for the TESLA cavity's 1.3 GHz TM$_{010}$ mode, normalized to $W_{{\rm total}}=1$~J of stored energy. The electric fields are axially symmetric about the $z$-axis.}
    \label{fig:cavity_Efield}
\end{figure}

For the TLS mechanism the loss tangent depends on the value of the electric field at a particular point $\bm{r}$ in the device. Finite element software COMSOL was used to predict the electric field distribution of the 1.3~GHz TM$_{010}$ mode for the TESLA cavity used in \cite{Romanenko2017}. The results are shown in 
Fig.~\ref{fig:cavity_Efield}, where it becomes evident that the value of electric field varies by several orders of magnitude at the internal surface of the cavity, ranging from zero at the top of the elliptical cell to $\approx 2 E_{{\rm acc}}$ at its edge. The ``accelerator field'' $E_{{\rm acc}}$ is defined as the accelerating voltage divided by the active cavity length \cite{Aune2000}.

As a result of this wide variation of electric fields, it is crucial to express $1/Q$ as an integral over the surface $S$ of the cavity; \Ch{it is also crucial to introduce an expression that interpolates between noninteracting (\ref{conv_tan}) and interacting (\ref{int_tan}) models for TLS saturation:
\begin{widetext}
\begin{equation}
   \frac{1}{Q} = \frac{1}{W_{{\rm total}}}\sum_{j}c_j \int_S |\bm{E}(\bm{r})|^2
   \left[\frac{1}{\ln{(\xi_j)}}\left(1-\frac{1}{\xi_j}\right)\ln{\left(\xi_j\sqrt{\frac{1+\left(\frac{E}{\xi_j E_{cj}}\right)^{2}}{1+\left(\frac{E}{E_{cj}}\right)^{2}}}\right)}+\frac{1}{\xi_j}\frac{1}{\sqrt{1+\left(\frac{E}{\xi_j E_{cj}}\right)^{2}}}\right]
   d^2r + \frac{1}{Q_{{\rm non-TLS}}}.
\label{Q_model}
\end{equation}
\end{widetext}}
Here $W_{{\rm total}}$ is the total electric energy inside the cavity, 
\begin{equation}
    W_{{\rm total}}=\frac{1}{4}\int_V \epsilon(\bm{r})\left|\bm{E}(\bm{r})\right|^{2}d^3r,
\end{equation}
and $c_j$ models the loss tangent arising from \emph{one particular ``$j$'' species of TLS}. This is given by
\begin{equation}
    c_j = \frac{\pi}{12}p_{j}^{2} \tanh{\left(\frac{\hbar\Omega_0}{2k_BT}\right)}\rho'_{j},
    \label{eqnc_j}
\end{equation}
where $p_j$ is its electric dipole moment and $\rho'_{j}$ is its energy-area density (dimensions of $({\rm energy}\times {\rm area})^{-1}$). The quantity $1/Q_{{\rm non-TLS}}$ models energy dissipation due to other mechanisms that do not saturate such as residual normal-state resistance due to thermal quasiparticles \cite{Mattis1958}, piezoelectric effect \cite{Diniz2020}, etc. 

\Ch{The square brackets of Eq.~(\ref{Q_model}) is chosen to satisfy the following properties for spectral diffusion parameter $\xi_j\geq~1$:

(1) $[\;]\rightarrow 1$ when $E\ll E_{cj}$;

(2) $[\;]\rightarrow \frac{1}{\ln{(\xi_j)}}\ln{\left(\frac{\xi_j E_{cj}}{E}\right)}$ when $E_{cj} \ll E \ll \xi_j E_{cj}$;

(3) $[\;]\rightarrow  \frac{E_{cj}}{E}\ll 1$ when $E\gg \xi_j E_{cj}$.

Therefore, Eq.~(\ref{Q_model}) interpolates smoothly between noninteracting and interacting loss tangents, in a fashion consistent with the asymptotic predictions of \cite{Faoro2012}}.

\section{Fitting experimental data}

Experimental data~\cite{Romanenko2017} for $Q$ as a function of $E_{{\rm acc}}$ at $T=1.5$~K for two different TESLA cavities is considered: (1) Electropolished cavity which was treated to remove most of the oxide layer on top of Nb, leading to a \Ch{thin $5$~nm oxide expected to have $0.5-1$~nm of NbO$_x$ on top of Nb followed by $\approx 4$~nm of crystalline Nb$_2$O$_5$ \cite{Altoe2022}; (2) Anodized cavity which contained a thick $100$~nm oxide layer, expected to be $0.5-1$~nm of NbO$_x$ followed by thick crystalline Nb$_2$O$_5$}. 

The $Q$ vs. $E_{\rm acc}$ experimental data points were extracted \cite{DataThief} from Fig.~2 (electropolished) and Fig.~4 (anodized) of \cite{Romanenko2017}. 
Only data points from the lowest $E_{{\rm acc}}$ up to $E_{{\rm acc}}= 1$~MV/m are included in our analysis since with larger $E_{{\rm acc}}$ the $Q$ starts decreasing with increasing $E_{{\rm acc}}$, signaling that an additional mechanism of loss takes over.

\begin{figure*}
\begin{center}
\subfloat[]{\includegraphics[width=.45\linewidth]{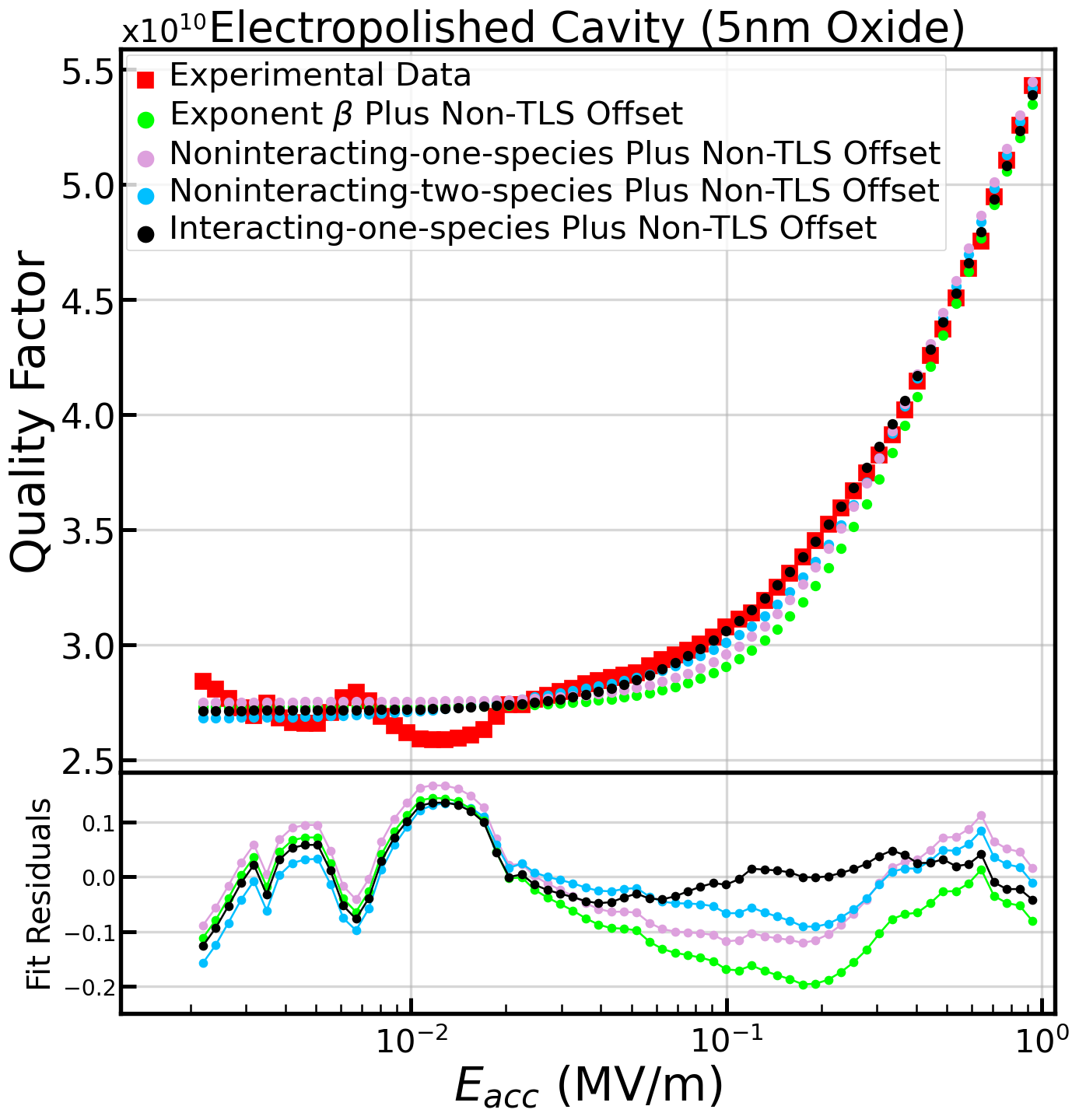}\label{fig:electropolished}}
\hspace{.45cm}
\subfloat[]{\includegraphics[width=.45\linewidth]{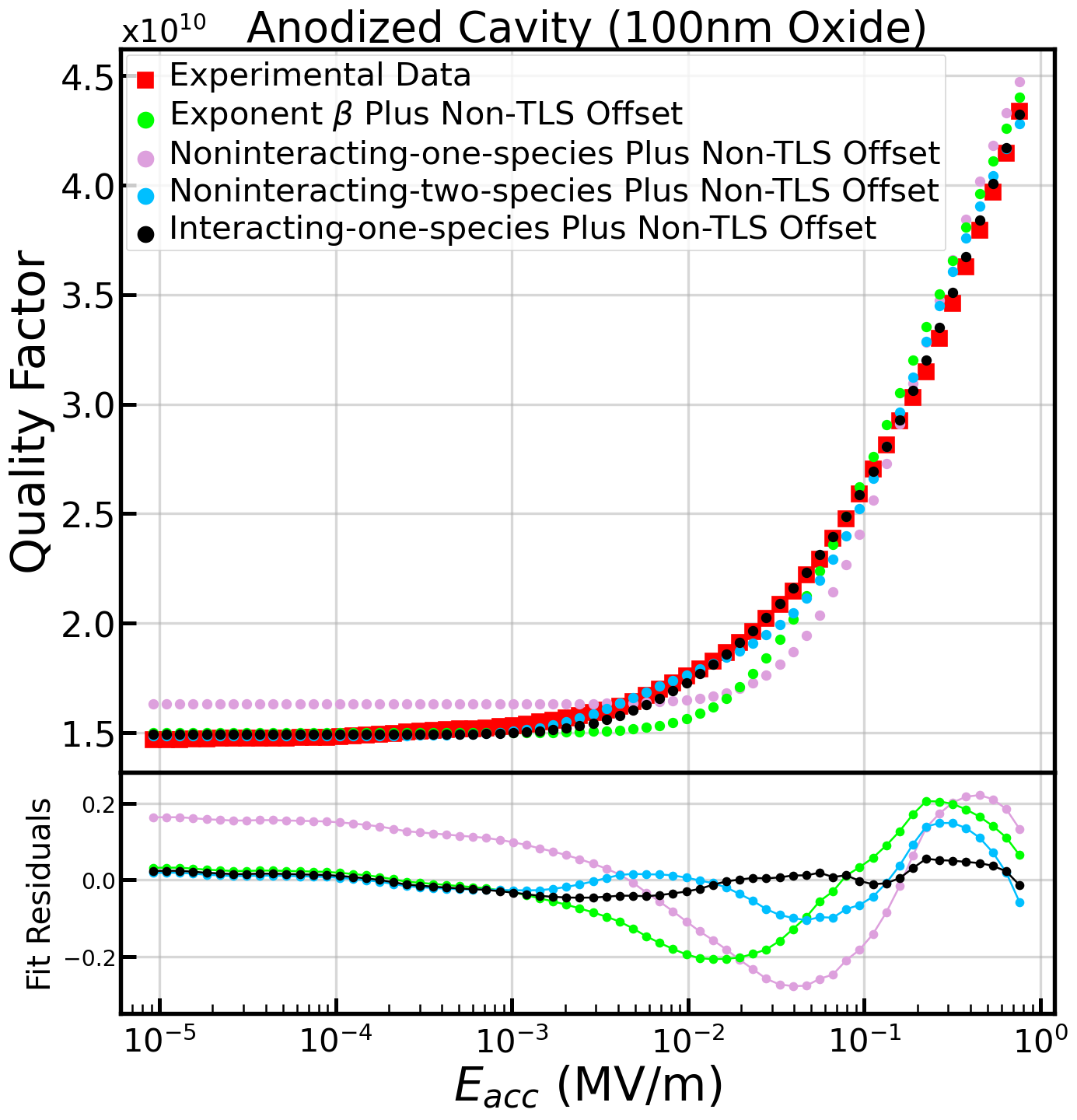}\label{fig:anodized}}
\end{center}
\caption{Quality factor $Q$ as a function of $E_{{\rm acc}}$ for the TM$_{010}$ mode of the TESLA cavity. The TLS model fit used in Ref.~\cite{Romanenko2017} (exponent $\beta$, Eq.~(\ref{romanenko_fit})) is shown as well. (a) Electropolished cavity, with $5$~nm thin oxide. Here the interacting-one-species-TLS plus non-TLS offset model provided the best fit with $\chi^2/{\rm DoF}=1.05$. The noninteracting-TLS plus non-TLS offset model led to $\chi^2/{\rm DoF}=2.56$ and $1.54$, for one species and two species, respectively.
(b) Anodized cavity, with $100$~nm thick oxide. Again the interacting-one-species-TLS plus non-TLS offset model yielded the best fit with $\chi^2/{\rm DoF}=1.01$. The fits for noninteracting TLSs led to $\chi^2/{\rm DoF}=24.5$ and $2.68$ for one and two species, respectively.}
\end{figure*}

The values of electric field at the surface of the cavity obtained by COMSOL for different $E_{{\rm acc}}$ were then used in conjunction with our Eq.~(\ref{Q_model}) to obtain best fits as a function of fitting parameters $c_j$, $\xi_j$, $E_{cj}$, and $Q_{{\rm non-TLS}}$. The oxide dielectric constant was assumed to be $\epsilon=33 \epsilon_0$ but this choice did not affect the fittings due to the small volume of the oxide. The results are shown in Fig.~\ref{fig:electropolished} (electropolished) and Fig.~\ref{fig:anodized} (anodized).

Various distributions and number of species were fit to the two experimental data curves using a nonlinear $\chi^2$ minimization algorithm. Uncertainty for the experimental error in measurements of $Q$ were estimated to be lower than 10\% in \cite{Romanenko2017}, but in our view this estimate encapsulates both the statistical and systematic uncertainties. Consequently, this value greatly overestimates the uncertainty required for the $\chi^2$ calculation. 

In order to calculate values of $\chi^2$ that properly represent fit quality and are able to detect overfitting, we obtained the statistical error $\sigma_{{\rm exp}}$ in the usual $\chi^2$ formula by taking the standard deviation of the $Q$ fluctuations in the low $E_{{\rm acc}}$ plateau region, where $Q$ is found to be independent of $E_{{\rm acc}}$. 
This led to $\sigma_{{\rm exp}} = 6.26 \times 10^{8}$ and $3.23 \times 10^{8}$ for the electropolished and anodized samples, respectively. 
The  $1\sigma$ errors for fitting parameters where found by manually adjusting one of the fitting parameters while holding the others constant until the value of $\chi^{2} /{\rm DoF}$ increased by $1$ (DoF is the number of degrees of freedom, equal to the number of data points minus the number of fitting parameters).

\Ch{The best fit for the electropolished cavity ($5$~nm thin oxide) was a \emph{interacting-one-species-TLS} plus non-TLS offset model. As seen in Table~\ref{table:fitting1}, this ``one-species'' fit led to a $\chi^2/{\rm DoF}=1.05$ quite close to $1$, indicating nearly optimal fit within experimental uncertainty. In contrast, the noninteracting-TLS models based on one and two species led to  $\chi^2/{\rm DoF}=2.56$ and $1.54$, respectively}.

We also included the fitting proposed in \cite{Romanenko2017} that was based on the expression 
\begin{equation}
  \frac{1}{Q} = \frac{F \delta_{TLS}(T)}{\left[ 1 + \left(\frac{E_{acc}}{E_c}\right)^{2} \right]^{\beta}} + \frac{1}{Q_{{\rm non-TLS}}}, 
\label{romanenko_fit}
\end{equation}
which lumped the electric field distribution inside the cavity into a single participation ratio $F$ for the oxide layer, and assumed the phenomenological exponent $\beta$ to be a free fitting parameter. For the electropolished cavity this led to \Ch{$\chi^2/{\rm DoF}=4.00$ and $\beta=0.38$, indicating a worse fit than our proposed interacting model.

The best fit for the anodized cavity ($100$~nm thick oxide) was also the interacting-one-species-TLS plus non-TLS offset, leading to $\chi^2/{\rm DoF}=1.01$. In comparison, the noninteracting model fits and the $\beta$ fit led to  considerably larger $\chi^2/{\rm DoF}$, as seen in Table~\ref{table:fitting1}}.

\begin{table*}[!ht]
  \resizebox{1.9\columnwidth}{4.cm}{
  \begin{tabular}{ccccc}
  \hhline{=====}
  
  
  \begin{tabular}{c}
    \textbf{Model}
    \vspace{-0.45cm}
  \end{tabular} &
  
    \multicolumn{2}{c}{\underline{\textbf{Electropolished Cavity (5~nm Oxide)}}}  &
    

    \multicolumn{2}{c}{\underline{\textbf{Anodized Cavity (100~nm Oxide)}}} 

  \\

  \begin{tabular}{c}
  \end{tabular} &

  \begin{tabular}{c}
    \boldsymbol{$\chi^2/{\rm DoF}$}
  \end{tabular} &

  \begin{tabular}{c}
    \textbf{Fitting Parameters}
  \end{tabular} &

  \begin{tabular}{c}
    \boldsymbol{$\chi^2/{\rm DoF}$}
  \end{tabular} &

  \begin{tabular}{c}
    \textbf{Fitting Parameters}
  \end{tabular} 

  \\

  \hline

  \begin{tabular}{c}
    Interacting One Species \\ Plus Non-TLS Offset
  \end{tabular} &
  \begin{tabular}{c}
    1.05
  \end{tabular} &

  \begin{tabular}{c}
    $E_{c} = (1.02 \pm 0.08 ) \times 10^5$~V/m\\
    $c = (6.04 \pm 0.16)  \times 10^{-24}$~C$^2$/J\\
    $\frac{1}{Q_{non-TLS}} = (1.19 \pm{0.04}) \times 10^{-11}$\\
    $\xi = 21.3^{+4.2}_{-2.9}$\\
  \end{tabular}& 
  
  \begin{tabular}{c}
    1.01
  \end{tabular} &

  \begin{tabular}{c}
    $E_{c} = (5.15 \pm 0.41 ) \times 10^3 $~V/m\\
    $c = (1.16 \pm 0.04)  \times 10^{-23}$~C$^2$/J\\
    $\frac{1}{Q_{non-TLS}} = (1.92 \pm{0.06}) \times 10^{-11}$\\
    $\xi = 205^{+32}_{-15}$\\
  \end{tabular}
  \\
  \hline

  \begin{tabular}{c}
    Two Species Plus \\Non-TLS Offset
  \end{tabular} &
  \begin{tabular}{c}
    1.54
  \end{tabular} &
  
  \begin{tabular}{c}
  $E_{c_1} = (2.42^{+14.9}_{-2.01}) \times 10^4$~ V/m\\
    $E_{c_2} = (3.29 \pm 0.28) \times 10^5$~V/m\\

    $c_1 = (6.30^{+5.45}_{-4.95} ) \times 10^{-25}$~C$^2$/J\\
    
    $c_2 = (5.45 \pm{0.26}) \times 10^{-24}$~ C$^2$/J\\
    
    $\frac{1}{Q_{non-TLS}} = (1.22 \pm{0.04}) \times 10^{-11}$\\ 
  \end{tabular}&

  \begin{tabular}{c}
    2.68
  \end{tabular} &

  \begin{tabular}{c}
    $E_{c_1} = (4.29 \pm{1.65}) \times 10^3$~V/m\\
    $E_{c_2} = (9.56 \pm{1.30}) \times 10^4$~V/m\\
    $c_1 = (3.70 \pm{0.70}) \times 10^{-24}$~C$^2$/J\\
    $c_2 = (7.81 \pm{0.41}) \times 10^{-24}$~C$^2$/J\\
    $\frac{1}{Q_{non-TLS}} = (1.98 \pm{0.05}) \times 10^{-11}$\\ 
  \end{tabular}

  \\ 
  
  \hline

  \begin{tabular}{c}
    One Species Plus \\Non-TLS Offset
  \end{tabular} &
  \begin{tabular}{c}
    2.56
  \end{tabular} &
  
  \begin{tabular}{c}
    $E_c = (2.95 \pm{0.27}) \times 10^{5}$~V/m\\
    $c = (5.81 \pm{0.25}) \times 10^{-24}$~C$^2$/J\\
    $\frac{1}{Q_{non-TLS}} = (1.22 \pm{0.03}) \times 10^{-11} $ \\
  \end{tabular}&

  \begin{tabular}{c}
    24.5
  \end{tabular}&

  \begin{tabular}{c}
    $E_c = (8.11 \pm{1.2} ) \times 10^{4}$~V/m\\
    $c = (1.04 \pm{0.04}) \times 10^{-23}$~C$^2$/J\\
    $\frac{1}{Q_{non-TLS}} = (1.84 \pm{0.03}) \times 10^{-11}$ \\
  \end{tabular}

  \\
  \hline

  \hline
  \begin{tabular}{c}
    Exponent $\beta$ Plus \\ Non-TLS Offset
  \end{tabular} &
  \begin{tabular}{c}
    4.00
  \end{tabular} &

  \begin{tabular}{c}
    $E_c = (1.90 \pm{0.17}) \times 10^5$~V/m\\
    $F\delta_{TLS} = (2.54 \pm{0.12}) \times 10^{-11}$\\
    $\frac{1}{Q_{non-TLS}} = (1.12 \pm{0.04}) \times 10^{-11}$\\
    $\beta = 0.38 \pm 0.05 $\\
  \end{tabular}& 

  \begin{tabular}{c}
    11.9
  \end{tabular} &

  \begin{tabular}{c}
    $E_c = (2.00 \pm{0.36}) \times 10^4$~V/m\\
    $F\delta_{TLS} = (5.25 \pm{0.38}) \times 10^{-11}$\\
    $\frac{1}{Q_{non-TLS}} = (1.42 \pm{0.08}) \times 10^{-11} $\\
    $\beta = 0.25 \pm{0.01}$\\
  \end{tabular}

  \\
 \hhline{=====}
 
\end{tabular}
}
\caption{Summary of (non)interacting one and two-species best fits with exponent $\beta$ fits shown for comparison.}
\label{table:fitting1}
\end{table*}

\section{Conclusions}

\Ch{In summary, an expression that interpolates between noninteracting and interacting models for TLS photon loss is proposed. When applied to experimental data in TESLA cavities, it shows that the best model fits are given by interacting TLSs with a sharp distribution of model parameters. This provides evidence that the TLSs present in niobium oxide are ``atomic-like'' (instead of ``glass-like'')}. 

\Ch{The fits obtained with our proposed expression (\ref{Q_model}) led to $\chi^2/{\rm DoF}$ quite close to $1$, that is $4-10\times $ lower than best fits using the phenomenological model with exponent $\beta$ and filling factor $F$ (\ref{romanenko_fit}). This result suggests the common method to fit TLS loss \cite{Wang2009, Macha2010, Sage2011, Burnett2016, Romanenko2017, Verjauw2021, Altoe2022} has to be revised in order to yield information on TLS microscopic parameters. Table~\ref{table:fitting1} also shows that the noninteracting-two-species model for the $5$~nm oxide yields $\chi^2/{\rm DoF}$=1.54 which is only $50\%$ worse than the best fit interacting-one-species model with $\chi^2/{\rm DoF}=1.05$. However the two-species model uses one fit parameter more and shows large fit residuals around $E_{{\rm acc}}=0.2$~MV/m. The data therefore strongly favors the interacting-one-species model.

It should be remarked that while the $Q(E_{\rm acc})$ plot has little structure, its gradual increase over several orders of magnitude \emph{can not be fit with any power-law model}. Note how the noninteracting and $\beta$ models deviate from experimental data at $E_{{\rm acc}}\sim 0.1$~MV/m in   Figs.~\ref{fig:electropolished}, \ref{fig:anodized}.}

To test whether continuous distributions of TLS parameters can provide an even better fit for the data, additional models with Gaussian and exponential distributions of parameters $c$ and $E_c$ were also considered as shown in Table~\ref{table:otherfittings}. 
These models did not provide better fits, supporting the conclusion that TLSs at the niobium/niobium oxide interface have a narrow distribution of $c$ and $E_c$, indicating a sharp distribution of $p$ and $T_1 T_{2}$ for individual TLSs, and thus \emph{small or no variation in microscopic structure}, yielding evidence for atomic TLSs. 

\Ch{The typical value of electric dipole moment for an atomic TLS is $p\sim |e|$\AA$\sim 10^{-29}$~Cm. Using this with Eq.~(\ref{eqnEc}) yields $\sqrt{T_1T_2}\sim 10^{-10}$~s for the 5~nm oxide, and $3\times 10^{-9}$~s for the 100~nm oxide. Using Eq.~(\ref{eqnc_j}) and assuming $\rho'\approx \sigma_{{\rm TLS}}/(\pi\tilde{\Delta})$ where $\sigma_{{\rm TLS}}$ is area density and $\tilde{\Delta}/k_B\sim 3$~K is the spread in TLS asymmetry due to O strain \cite{Enss2005}, one gets $\sigma_{{\rm TLS}}\sim 1.4 \times 10^{11}$/cm$^{2}$ for the $5$~nm and $2.7 \times 10^{11}$/cm$^{2}$ for the $100$~nm oxide. 

The estimates above indicate the best fit parameters in Table~\ref{table:fitting1} are physical, e.g. one out of $\sim  10^{5}$ atoms act as a TLS for the $5$~nm oxide, $10\times$ less for the $100$~nm oxide because of its thick crystalline Nb$_2$O$_5$ \cite{Altoe2022}. However, it is not possible to distinguish between two possible scenarios: (1) That the dominant TLS in the different oxides are distinct microscopic species, or (2) that they are the same microscopic species with different area densities and embedded in a different environment. Just from $Q(E)$ data alone it is not possible to distinguish between these two scenarios.}

The best fit for the electropolished cavity implies the $E,T\rightarrow 0$ limit for the loss tangent in the $5$~nm oxide is  $\tan{(\delta_{{\rm TLS}})}=4c \times 2k_B\times 1.5 {\rm K} /(\epsilon \times 5~{\rm nm}\times h \times 1.3~{\rm GHz})=8\times 10^{-4}$. This value is 13$\times$ smaller than the estimate that neglected the electric field distribution \cite{Romanenko2017}, and is 4$\times$ smaller than the typical \Ch{surface/interface} loss tangent measured in quantum computing devices \cite{Kaiser2010, Wang2015}. The reduced $\tan{(\delta_{{\rm TLS}})}$ for TESLA cavities demonstrates the high quality of its oxide. 

%

\Ch{The proposed numerical modelling based on Eq.~(\ref{Q_model}) is general in that it applies to other materials and devices with multiple interfaces and substrates. 
Further experimental characterization based on e.g. electron microscopy and time-of-flight secondary ion mass spectrometry \cite{Dhakal2018, Murthy2022} \emph{in tandem} with dielectric loss measurements is required to identify the microscopic structure and atomic composition of the dominant TLSs.}

\begin{table*}[!t]
 \centering
 \resizebox{1.8\columnwidth}{2.5cm}{
 \begin{tabular}{ccc}
  
 \hhline{===}
  
  \begin{tabular}{c}
  \textbf{Model}
  \end{tabular} &
  
  \begin{tabular}{c}
  \underline{\textbf{{Electropolished Cavity (5~nm Oxide)}}} \\ \boldsymbol{$\chi^2/{\rm DoF}$}
  \end{tabular} &

  \begin{tabular}{c}
  \underline{\textbf{Anodized Cavity (100~nm Oxide)}} \\
  \boldsymbol{$\chi^2/{\rm DoF}$}
  \end{tabular}
  
  \\
  \hhline{===}
  
  \begin{tabular}{c}
    Gaussian $E_c$ Plus \\ Non-TLS Offset
  \end{tabular} &
  
  \begin{tabular}{c}
    3.24
  \end{tabular}&

  \begin{tabular}{c}
    26.3
  \end{tabular} 

\\

   \hline
  \begin{tabular}{c}
    Gaussian Electric Dipole \\ Plus Non-TLS Offset
  \end{tabular} &
  \begin{tabular}{c}
    3.36
  \end{tabular}&

  \begin{tabular}{c}
    29.9
  \end{tabular} 

\\

    \hline
  \begin{tabular}{c}
    Exponential Electric Dipole \\ Plus Non-TLS Offset
  \end{tabular} &
  \begin{tabular}{c}
    4.32
  \end{tabular}& 

  \begin{tabular}{c}
    29.3
  \end{tabular} 

 \\
   \hline
  \begin{tabular}{c}
    Exponential $E_c$ Plus \\ Non-TLS Offset
  \end{tabular} &
  \begin{tabular}{c}
    4.43
  \end{tabular}& 

  \begin{tabular}{c}
    12.3
  \end{tabular}

 \\
  
      \hline
  
  \begin{tabular}{c}
    Exponential $\frac{1}{f}$ Noise \\Plus Non-TLS Offset
  \end{tabular} &
  \begin{tabular}{c}
    47.5
  \end{tabular}&

  \begin{tabular}{c}
    27.9
  \end{tabular} 

\\
 \hhline{===}
 
\end{tabular}
}
\caption{Summary of additional noninteracting fits that assume a continuous distribution of parameters $c$ and $E_c$.}
\label{table:otherfittings}
\end{table*}

\begin{acknowledgments}
The authors thank A. Blackburn and R. McFadden for useful discussions and insight into various topics. This work was supported by NSERC (Canada) through its Discovery (Grant number RGPIN-2020-04328), and Undergraduate Student Research Award programs.
\end{acknowledgments}

\newpage


\input{two_species.bbl}

\end{document}

%% file: two_species.bbl
%